**The limits of selection under plant domestication**

Robin G Allaby[1*], Dorian Q Fuller[2], James L Kitchen[1,3]

1. School of Life Sciences, Gibbet Hill Campus, University of Warwick, Coventry CV4 7AL
2. Institute of Archaeology, University College London, 31-34 Gordon Square, London WC1H 0PY
3. Computational and Systems Biology, Rothamsted, Harpenden, Herts.

* corresponding author email: r.g.allaby@warwick.ac.uk

**Abstract**

Plant domestication involved a process of selection through human agency of a series of traits collectively termed the domestication syndrome. Current debate concerns the pace at which domesticated plants emerged from cultivated wild populations and how many genes were involved. Here we present simulations that test how many genes could have been involved by considering the cost of selection. We demonstrate the selection load that can be endured by populations increases with decreasing selection coefficients and greater numbers of loci down to values of about $s = 0.005$, causing a driving force that increases the number of loci under selection. As the number of loci under selection increases, an effect of co-selection increases resulting in individual unlinked loci being fixed more rapidly in out-crossing populations, representing a second driving force to increase the number of loci under selection. In inbreeding systems co-selection results in interference and reduced rates of fixation but does not reduce the size of the selection load that can be endured. These driving forces result in an optimum pace of genome evolution in which 50-100 loci are the most that could be under selection in a cultivation regime. Furthermore, the simulations do not preclude the existence of selective sweeps but demonstrate that they come at a cost of the selection load that can be endured and consequently a reduction of the capacity of plants to adapt to new environments, which may contribute to the explanation of why selective sweeps have been so rarely detected in genome studies.

**Introduction**

Domestication is an evolutionary process that provides a cornerstone to understanding the mechanism of selection (Darwin 1859). In the case of plants the evolution of domestication involves the selection of a characteristic group of traits that are collectively termed the domestication syndrome (Harlan *et al.* 1973, Hammer 1984). These traits include the loss of shattering, changes in seed size, loss of photoperiod sensitivity and changes in plant and spikelet architecture (Fuller 2007). It is interesting that among crops of a similar type such as cereals, the number of syndrome traits is also similar. It is not known whether this represents the maximum number of traits that could have been selected within the time period of syndrome fixation spanning several thousand years, or whether further traits could have been selected in each case but were not. There has been much debate about how these traits were selected, the pace and strength of selection, and the extent to which evolution under domestication continues today (Fuller 2007, Brown *et al.* 2009, Honne & Heun 2009, Purugganan and Fuller 2009, Allaby *et al.* 2010, Abbo *et al.* 2010, 2011, Fuller *et al.* 2011).

Classic field trials of experimental harvesting of wild progenitors of wheat suggested that in the case of cereals selection coefficients as high as 0.6 could have been in operation during domestication resulting in fixation of loss of shattering traits within a few decades (Hillman And Davies 1990). These findings have been supportive of a rapid transition model of agricultural origins (Diamond 2002) in which domesticated forms of crops appeared over a very short time period in a 'Neolithic Revolution' (Zohary and Hopf 2000). However, an emergent feature of the archaeological record in recent years has been the protracted appearance of domesticated crops (Asouti and Fuller 2013, Tanno and Willcox 2006, 2012, Weiss *et* *al.* 2006, Willcox 2005, Willcox *et al.* 2008, Willcox and Stordeur 2012, Hillman et al 2001) and estimates of selection coefficients made directly from the archaeobotanical record have been as low as 0.003 (Purugganan and Fuller 2011, Fuller *et al.* 2014). Under a protracted scenario of domestication, the expected patterns of genetic diversity need to be re-evaluated in order to interpret the evolutionary history of domestication (Allaby *et al.* 2008, Allaby 2010). For instance, under protraction traits may have been selected more slowly in the face of gene flow between cultivated and wild populations resulting in the appearance of relatively weak selection coefficients. This extended time period would increase the opportunities for parallelisms in syndrome traits as similar traits are independently selected in distinct geographic locations, and genomic mosaicism associated with phylogeography could result (Allaby 2010). It could also be the case that the protracted process allowed more traits to be selected than would have been possible under the restricted time period of a rapid transition of a few tens to hundreds of years. It is therefore useful to establish how much selection would have been possible to drive the evolution of the domestication syndrome under protracted and rapid transition scenarios.

Selection comes at a cost in that some organisms must die before reproducing each generation for their genes to be selected against, so causing a reduction in the overall population size. A consequence of this cost is that the amount of selection that a species can withstand is limited. Haldane noted that for this reason there is a limit to the number of traits that plant breeders are able to select at a given time, and that the pace at which evolution can be driven by natural selection is limited because of the cost of selection (Haldane 1957). In order to understand the pace of selection of the domestication syndrome traits it is necessary to know the number of genetic loci involved. Increasingly, the underlying genetic bases of the domestication syndrome are being elucidated (Fuller and Allaby 2009), and the number of loci associated with their control has recently been estimated to be 27, and as much as 70 in tetraploid wheats (Peleg *et al*. 2011, Peng *et al*. 2003). Estimates of loci under selection from the genome analysis of other crops have yielded a range of numbers, with up to 1200 loci suggested in maize, but an expectation of around 40 at the signal strength of *tb1* (Wright *et al.* 2005) and 36 loci in sunflower (Chapman *et al.* 2008). The power to detect signatures of selection in genetic data is limited (Yi et al 2010), so it is unclear from these studies whether these numbers of loci under selection represent the totality of loci or simply the most strongly selected loci that reach the threshold of detectability. These numbers are high relative to the typical number of syndrome traits because some traits, such as seed size, are under polygenic control (Gupta *et al.* 2006), while other traits, such as loss of shattering, are under the control of one or just a few genes (Konishi *et al.* 2006, Li *et al.* 2006, Li & Gill 2006, Takahashi *et al.* 1955, Ishikawa et al 2010; Ishii et al 2013). Intuitively, it seems likely that traits under monogenic control may be subject to stronger selection pressures than those under polygenic control because the selection coefficients associated with each locus for a trait have an additive effect. Therefore, as loci governing a trait are progressively added, the value of *s* for each locus must progressively reduce in order to maintain the same overall selection pressure on the individual organism. Consequently, it might be expected that those traits of the domestication syndrome that are under monogenic control would have been under the strongest selection and appeared the earliest. Surprisingly, the reverse is observed in the archaeological record in that the tough rachis mutant appears to be selected slowly and late relative to other traits (Tanno and Willcox 2006, Fuller 2007), and increase in seed size appears very early on in the archaeological record despite the complexity of its genetic control and a rate of selection that is not significantly less than for shattering (Purugganan and Fuller 2011, Fuller *et al.* 2014). While the timing of the first appearance of traits is attributable to a sequence of different behaviors of proto-farmers subjecting different selection pressures on plants at different times (Fuller *et al.* 2010, 2011), the surprising similarity in rates highlights the need to better understand the selection pressures involved in domestication.

In this study the limitations imposed by the cost of selection were examined through computer simulations to establish the relationship between the number of loci under selection, the strength of selection and the ability of plant populations to recover from reduced population sizes resulting from rounds of selection. The simulations were executed under a scenario based on the archaeological record in which traits appear and are selected over a period of 3000 years (Fuller 2007, Fuller *et al.* 2014, Tanno and Willcox 2006, 2012). Populations of virtual plants were endowed with a number of loci, which were considered unlinked to each other and inherited through a process of random segregation. In reality the overwhelming majority of domestication syndrome loci have been identified to be regulatory in nature (Purugganan and Fuller 2009, Meyer and Purugganan 2013). Consequently, it can be inferred that most mutations have been epistatic in their effect. One can consider the changes in gene expression caused by mutation at a distal regulatory locus to be a phenotypic consequence, and that the focus of selection would act on the regulatory locus rather than the regulated locus. Such a situation can be reasonably modeled by treating each locus under selection as independent to other loci. This model makes no assumption about the function of the loci under selection, and so the stage of the life cycle in which the resultant phenotype is expressed, and is inclusive of epistatic mutations. Furthermore, phenotypic traits may be under monogenic or polygenic control. A model of independent selection of loci inherently describes traits under monogenic control. Traits under polygenic loci are also described if it is assumed that the overall trait, such as seed size, is contributed to by independently segregating loci, which is largely true for quantitative traits. One might also consider that the selective value of a mutation at a locus is dependent on the presence of mutations at other loci. However, this would require a specific knowledge of the dependencies, which would preclude a general model. Finally, it is known that some domestication syndrome loci are located in close proximity to each other (Gepts 2004), and that positively selected mutations may be associated with either linked deleterious loci, or have slightly deleterious pleiotropic effects (Bomblies and Doebley 2006). In the case of tightly linked loci, it is reasonable to model a single locus and the associated selection coefficient thereby represents an overall selective value. The general effect of linkage disequilibrium can reasonably be explored with a model of independently segregating loci by including both inbreeding and outbreeding mating systems in which linkage disequilibrium will be high and low respectively.

Given the assumptions outlined, it is reasonable to use a general model of mutation selection of independently segregating loci in order to assess the amount of selection that a plant population can endure under domestication. Each locus was associated with a selection coefficient ($s$), which selected against the wild type allele. Mutants were generated for loci in the population that had a fitness value of 1, so they were not selected against. Individuals survived with a probability equal to the product of the fitness values of alleles across all loci. Under this system, the resulting individuals in the next generation would be fewer than the previous generation. The ability of the population to recover from such a round of selection was determined by a maximum fecundity parameter where each individual was capable of having more than one progeny, causing the population to expand. However, population expansion was tempered by both environmental checks and an environmental carrying capacity. Therefore the number of individuals generated in the next generation was expanded from the current generation value by an amount determined by the maximum fecundity parameter (*mf*) up to the carrying capacity population size that could not be exceeded. For each set of simulation conditions the probability of extinction, severity of selection bottleneck, and the rate and extent of fixation of domestication syndrome trait controlling alleles were determined.

It is well known that the rate of fixation of a mutation may be slowed down by gene flow from an adjacent environment in which the same selection regime does not apply. This is true in the case of the domestication syndrome in which the wild and cultivated environments provide diametrically opposed selective forces on several traits (Allaby 2010). The consequence of such gene flow is to reduce the proportion of the advantageous mutant so in effect to reduce its selective advantage, and therefore also the resulting selection coefficient between the mutant and the wild type. In an explicit model gene flow would therefore effectively reduce the selection coefficient from the input value making it an unknown parameter, which is undesirable in this case. The cost of selection is correspondingly reduced as each wild type individual in the local population that fails to reproduce is offset to some extent by immigrant wild types. It is therefore a reasonable simplification to exclude gene flow from the model in this case and consider the selection coefficients applied as comparable to the resultant selection coefficient in the face of gene flow in terms of how the cost of selection is limiting. The replenishment of the population through migration-mediated gene flow (as opposed to purely pollen flow) is assumed to be negligible, and the prevention of population extinction by the continual arrival of a few individuals not carrying adaptive mutations is a largely uninformative parameter.

**Results and Discussion**

***The maximum number of loci under selection.*** In the first set of simulation experiments a model in which increasing numbers of loci subject to the same selection coefficient was applied. Two regimes were considered relating to the ability of members of the population to reproduce. The first was a conservative regime in which there was an underlying assumption that the organisms were held in close check by organisms of different species in the environment, close to Darwin's original insight that the typically geometric potential for species to reproduce is held back by complex interspecific competition. Under this regime the *mf* parameter was set to 1.5, such that a population was capable of expanding 50% at most per generation. The second regime was considerably more liberal and *mf* was set to 10, allowing a ten-fold expansion. The underlying rationale in this case was that a cultivated environment inherently reduces interspecific competition, excepting that which is due to human predation. The extent to which a crop population can expand is a function of the number of propagules generated per plant, and the proportion of the harvest that is set aside for sowing the following year. Previous studies have suggested that a quarter of a crop harvest may be set aside representing about an eighth of the propagules generated in a generation as it was estimated that approximately 50% of seeds were harvested, with the rest lost due to dispersal and not sown (Hillman and Davies 1990). Under such a regime an eightfold increase per individual would maintain an overall constant population size, or an *mf* value of 1. However, cereals such as barley have up 20 or 30 grains per plant, which would lead to a 3-4 fold increase in population size per generation under these conditions. In reality it is likely that a lower proportion of the grain would be sown to compensate bringing the *mf* parameter to something close to our conservative regime, however we selected an *mf* value of 10 to explore the possible effect that could be introduced by human agency. We consider an *mf* of 10 to be in excess of what is likely to be achieved.

Simulations were carried out on populations that began with 1000 individuals to reflect a reasonable size based on genetic diversity studies (Zhu *et al.* 2007, Eyre-Walker & Gaut 2007), but it should also be remembered that the pace of selection for a given allele under a given selection coefficient is expected to be independent of population size (Haldane 1924). This should not be confused with the parameter of selection intensity ($2N_e s$) often used in coalescent approaches that describes the impact of selection on genetic diversity (Innan and Kim 2004), which *is* dependent on population size. Each set of parameters were repeated 100 times. The two fecundity regimes were explored using two mating systems, a 2% inbreeding population, similar to that of wheat or barley, and a 98% out-crossing population to represent out-crossing crops (tables S1-4). These mating systems represent the normal biases expected in plants (Vogler and Kalisz 2001). The simulation outputs were used to construct probability of survival landscapes under the four possible regimes, Figure 1. It is notable that in all simulation parameters explored, the switch from survival to extinction with increasing loci number under selection is precipitous, with populations going from an estimated probability of survival of one to zero with the addition of just 1-5 loci in most cases. This change is more abrupt in out-crossing populations than inbreeding ones. Generally, very few loci could be simultaneously under selection at selection coefficients of 0.1 or higher, and values of *s* higher than 0.3 were unlikely to be survived by populations under the conservative *mf* regime. We found that 80 and 83 were the maximum number of loci that could be selected under the conservative expansion regime before extinction occurred under the lowest value of *s* explored (0.005). This value is close to typical values under natural selection, and that calculated from cereals in the archaeological record (Purugganan and Fuller 2011, Fuller *et al.* 2014). Under the higher *mf* regime we found that 227 and 230 loci could be under selection in out-crossing and inbreeding systems respectively at the lowest value of *s* explored (0.01).

The probability of survival became less than one in populations that generally experienced bottlenecks of less than 30% and less than 10% at the lower values of *s*. While bottlenecks of this extremity may be found in nature, it is unlikely that such a bottleneck could be tolerated under a cultivation regime since the majority of the food source would disappear. We constructed landscapes to view the effect of selection on bottleneck size, Figure 2. A precipitous drop is still discernible under the conserved *mf* regime, which occurs at a threshold point when the bottleneck size is between 60 and 70% of the initial population size. In the case of the higher *mf* regime there is a steady decline in bottleneck size. We judge that a reasonable level of bottleneck that could be tolerated by cultivators is 60-70% before there is too great a reduction in food production for it to be worthwhile investing in cultivation. Interestingly, all regimes explored suggest that the number of loci that could be under selection for such a bottleneck lie within the range of 50-100 loci (figure S1). This result is particularly interesting because it closely mirrors the number of loci under selection identified in genome studies of major crops (Peleg *et al.* 2011, Peng *et al.* 2003, Wright *et al.* 2005, Chapman *et al.* 2008).

***Selection from standing variation.*** The simulations so far consider mutants that occur in very low frequencies in wild populations that are generally selected against in the wild. A number of regulatory genes have been identified in maize that have been selected during domestication which occur in a wide range of frequencies (0.1-0.88) in the wild progenitor (Weber et al 2007, Studer et al 2011). In these sorts of cases it is not expected that the wild and cultivated populations are subject to diametrically opposed selection regimes, but instead aspects of adaptations to the wild environment or neutral standing variation have selective value in the cultivated environment. Models have shown that even strong selection will likely not leave a detectable signature of selection from standing variation at higher frequencies (Innan and Kim 2004, Teshima et al 2006) because less of the adjacent genomic variation is lost during the sweep process. Consequently, this part of the selection process during domestication would be largely invisible from genome diversity scan approaches. To investigate whether our estimates of the number of loci that could be under selection were different from standing variation rather than spontaneous mutation, we carried out simulations in which the starting frequency of the mutant was 0.5, Figure S2. Generally, populations could sustain a larger number of loci under selection from standing variation. Under these conditions there was a greater difference between the mating strategies than for selection from spontaneous mutation. Outbreeding populations sustained about 50% more loci before population extinction was observed, whereas inbreeding populations sustained over 100% more loci. The difference between the mating strategies is likely explained by the difference in heterozygote proportions between the two population types. Inbreeding populations hold most of the recessive mutations in homozygous individuals, which therefore confer an advantage to the individual. The difference between selection from standing variation and spontaneous mutation is less pronounced when the population bottleneck is considered, Figure S3. The 60-70% threshold that we consider to be a realistic pragmatic limit for cultivation is reached in the 50-110 loci range from standing variation for weak selection (s = 0.005).

***Co-selection and the interference between loci.*** In reality it is unlikely that all loci would be under equal selection pressure. Therefore in the third set of experiments we explored mixed selection regimes in which there was one strong selection pressure, combined with varying numbers of weakly selected loci. In this case we selected a value of 0.3 for *s* to represent strong selection, and 0.01 for *s* to represent relatively weak selection. In this set of experiments we only considered the conservative fecundity regime in which *mf* was set at 1.5.

In a mixed regime the maximum number of weak loci that could be added to the strongly selected loci before extinctions occurred was 6 and 4 for the inbreeding and out-crossing regimes respectively (table S5). The total amount of selection that a population is subject to can be expressed by a 'selection load', which we define as the sum of the selection coefficients of the loci under selection. In this mixed regime a selection load of 0.34-0.36 was endured by populations, which is less than that endured by populations subjected to selection at *s* set to 0.01 only (0.4-0.41). A notable effect was that rate at which advantageous mutants of loci subject to strong selection were fixed increased in the presence of selection of increasing numbers of weakly selected loci, but the effect is not clearly apparent in the weakly selected loci, Figure 3. Since loci were unlinked, the rate increase is unlikely to be due to a hitchhiking effect. Previously, it has been suggested that an additive selective advantage effect could occur in individuals that received mutants with adaptive values to two unrelated selection regimes from different source populations, resulting in a type of cultivation magnetism between the two source populations (Allaby 2010). An alternative explanation that might be forwarded is that the reduced population size in the bottleneck caused by the increased selection load of many weakly selected loci may increase the rate at which the strongly selected loci are fixed, although we do not expect this to be the case from previous theory (Haldane 1924). To distinguish between these possibilities we carried out a set of simulations on just the one strongly selected locus in smaller populations that closely matched bottleneck sizes in the mixed selection regimes. In the case of the out-crossing population a size of 500 individuals was selected, which closely matches the bottleneck of the mixed regime with 5 weakly selected loci, and a size of 300 was selected for the inbreeding population, which closely matches the mixed regime with 10 weakly selected loci. In both cases the smaller population size did not produce fixation rates as fast as with the mixed regime, showing that the additive effect of loci is a more likely cause than reduced population size.

The interference in the rate of selection caused by co-selection was investigated further by plotting the rate of fixation of increasing numbers of loci under uniform regimes of either strong or weak selection, Figure 4. Under these conditions the additive advantage occurred in out-crossing populations, but increased loci burden led to a decrease in the rate of fixation of inbreeding populations. This result can be explained by the mating strategy. The rate at which advantageous mutations are united in single individuals by crossing is lower in an inbreeding system than in an out-crossing one. Consequently, in an inbreeding system although an individual may have a single advantageous mutation, it will have numerous disadvantageous alleles at other loci that will depress its probability of success. The result is that the numerous loci will interfere with each other to an extent dependent on the number of loci involved. This effect of mating system is similar to the Hill-Robertson effect noted in *Oryza* (Hill-Robertson 1974, Flowers *et al.* 2012) in which diversity is depressed in inbreeders, although the loci are physically unlinked in this model demonstrating that mating system alone is highly influential as well as genome location for interference of selection of loci. The exception in this study is when a strongly selected locus is in the presence of weakly selected loci in an inbreeding population. In this case it appears that the depressing effects of the weak loci have little effect on the strongly selected locus, but the locus does benefit from the additive effect of recruiting weakly advantageous mutations.

The out-crossing populations show a less complex pattern in co-selection effects. In both weakly and strongly selected loci, an increase in loci number increases the rate at which each locus is fixed. In the case of weakly selected loci, the number of loci has to be relatively large before the effect becomes clear. The effects of co-selection suggest that any particular trait will be selected for more quickly in the presence of a second, unrelated, selection pressure acting on an unlinked locus. Consequently, there appears to be a synergistic consequence to multiple unrelated selection pressures that could drive an organism to adapt to new environments and outcompete organisms adapting to fewer environments.

***Selection load and an optimum pace of selection at the genomic level.*** The total selection load that populations were subjected to without extinctions occurring increased with decreasing selection pressure per locus, Table S6. This leads to an interesting dynamic, because although more strongly selected loci are fixed more quickly, the total amount of selection that can be endured at one time is less, which is reminiscent of the tale of the tortoise and the hare. This demonstrates that an adaptive regime over many loci can endure an environment which is more complex and overall harsher than one in which a single mutant is acutely selected, therefore rendering such complex harsher environments open to the 'tortoise' which would be unavailable to the 'hare'. Hares, on the other hand may be better suited to adapt acutely to extreme environments of low complexity. It is also notable that populations under strong selection pressures per locus are vulnerable to extinction at bottleneck sizes that are considerably larger than in the case of weak selection pressures, for instance between 40 and 50% in the mixed selection regime (table S5), but only 14-15% under comparable weak selection (tables S1 and S2). The hare strategy therefore is much more risky, achieves less selection overall due to the risk of extinction and as a consequence would likely produce populations that were ultimately weaker.

Ever decreasing levels of selection per locus will eventually lead to no selection at all, which raises the question is there an optimum number of loci and level of selection? The extent to which the survivable selection load increases with decreasing selection per locus diminishes with the latter parameter (Figure S4). In the case of the conservative reproductive regime where the *mf* parameter was set to 1.5, the selection load would not be expected to be much greater than 0.41 observed at a level of selection at which *s* is set to 0.005. Similarly, in our liberal reproductive regime there may be a little improvement on the selective load of 2.3 observed by reducing *s* from 0.01 to 0.005. Therefore the advantage of an increased selective load is not likely to be greatly improved at selection coefficients much below 0.005. It is perhaps for this reason that the selection coefficients observed in nature tend to be of this order of magnitude, and perhaps also one reason why apparently disparate selection regimes on different domestication syndrome traits have yielded very similar measurements of selection coefficients, also in this order of magnitude (Purugganan and Fuller 2011; Fuller et al 2014).

We finally calculated the average maximum rate of substitution achieved under each of our selection regimes without extinctions occurring (Figure S5). Generally, a larger number of substitutions had been achieved per generation under the lower selection regimes by the time all mutants would have been fixed. Haldane originally estimated that one substitution every 300 generations corresponding to a selection coefficient of 0.1 was about what could be expected of natural selection (Haldane 1957). Our models agree quite well with this estimation, and suggest that at lower levels of selection across the genome, the pace could be considerably faster.

***The pace of selection of the domestication syndrome in plants.*** The simulations in this study demonstrate that regardless of the reproductive capacity of plants there is an optimum level of selection across a plant genome in the order of *s* equal to 0.005 that should apply to natural selection as well as selection in the cultivated environment. Co-selection will tend to further push genome evolution towards a maximum number of loci under selection in out-crossing species. Under cultivation there is likely to be a stringent restriction on bottleneck size in order to maintain a viable food source, and we estimate that as a result the expectation of the number of loci under selection should be in the range of around 50-100 loci. This value is true of selection from spontaneous mutants as well as from standing variation. These findings suggest that genome wide efforts to detect signatures of selection in crops are probably recovering most of the loci under selection, and that those loci were most likely selected from spontaneous mutations since selection from standing variation is less likely to be detected in genomic signatures. The difference between the number of signatures of selection detected and the upper limits described here reflects the amount of selection that could have occurred from standing variation. This suggests that domesticated wheat was mostly based on spontaneous mutation, maize and sunflower may have had progressively more of their domestication adaptations from standing variation. Germane to this observation is that selection of recessive mutants is quicker in inbreeding populations where the majority of individuals are homozygous through selfing. Therefore, it would have been easier for the inbreeding crops such as wheat and barley to select spontaneous mutation, while maize and sunflower would likely have had a greater pressure to incorporate standing variation.

While these simulations do not preclude the existence of selective sweeps, they do show that sweeps come at a cost of reducing the selection load that a population is capable of enduring. This could explain why sweeps are rarely observed in nature, but also why agricultural expansion was repeatedly associated with collapse in new environments shortly after arrival (Shennan *et al.* 2013, Stevens and Fuller 2012). It is possible that the rapid pace of expansion could have forced equally rapid adaptation of plants to latitude, which would have required strong selection of a low number of loci – an adaptation of low complexity. Given the dynamic complex environment into which agriculture had advanced, it may have been the case the plant populations were incapable of further adaptation to changing conditions as they occurred. To better understand the expansion of agriculture further consideration is needed of the pace of movement across the latitudinal selection gradient in the context of tolerable limits of plants, and whether different paces are associated adaptations of low and high complexity respectively (Kitchen and Allaby 2013).

**Methods**

The simulations were carried out using a program written by R.G.A. For the program details and methodology, see SI Text. Simulations were carried with populations of 1000 individuals. The simulation begins with an initialization of the population in which hermaphrodite individuals are assigned wild type alleles for the defined number of loci under selection. A mutation rate of 0.001 was used, such that on average a single recessive advantageous mutation would appear each generation in populations that had no mutants. In subsequent generations, individuals were generated by randomly selecting gametes (themselves generated through a process of random segregation with all loci unlinked) from individuals of the previous generation with a probability of selecting the same donor twice equal to the mating strategy (0.02 for out-crossing simulations and 0.98 for inbreeding populations). Newly generated individuals then survived with a probability equal to the product of the fitness values of the alleles they carried, such that the probability of survival (*su*) was defined as:

$$su = \prod_{1}^{k} \omega_i \qquad (1)$$

For *k* loci, where $\omega_i$ is the fitness of the *i*th locus as given by:

$$\omega_i = 1 - s_i \qquad (2)$$

Where $s_i$ is the selection coefficient of the *i*th locus. The selection coefficient of the *i*th locus was moderated by the value lambda for heterozygotes ($S_{het}$) such that

$$s_{ihet} = (1 - \lambda)s_i \qquad (3)$$

A value of 0 was taken for lambda in all simulations in this study to represent recessive mutations, which represent the majority of known mutations associated with domestication. In the first generation this step was repeated for a number of times equal to the population size, and inevitably led to a number of individuals in the next generation which were fewer than this value. In subsequent generations the number of attempts at making new individuals was given by

$$N_{attempts} = N_{n-1} mf \qquad (4)$$

For ($N_{attempts}$ < initial population size), where $N_{attempts}$ is the number of individuals created then challenged, $N_{n-1}$ is the number of individuals in the previous generation and $mf$ is the maximum fecundity parameter. Where this condition was violated, the initial population size was used, representing the carrying capacity of the environment. This process was iterated for the specified number of generations.

Simulations were carried out for 3000 generations. Each set of simulation conditions was repeated 100 times, and average frequencies of advantageous mutants for each locus and population sizes were recorded for each generation. The number of generations per substitution was calculated by dividing the time to fixation by the number of loci that had been fixed under maximum selection load conditions for a given value of $s$. Time to fixation ($F$) was approximated when fixation had been incomplete at the end of simulations using a logistic sigmoid function equation (5).

$$F \approx \frac{13.8G}{6.9 - \ln\left(\frac{1-f}{f}\right)} \qquad (5)$$

Where $G$ is the number of generations in the simulation and $f$ is the final frequency of the mutant under selection.

**Acknowledgements**

This work was kindly supported by the Leverhulme Trust (grant number F/00 215/BC).

Figure 1. Probability landscapes of population survival ($p$) for a given number of loci ($g$) under selection coefficient ($s$). A. Inbreeding population with $mf$ of 1.5. B. Out-crossing population with $mf$ of 1.5. C. Inbreeding population with $mf$ of 10. D. Out-crossing population with $mf$ of 10.

Figure 2. Landscapes of minimum population bottleneck ($b$) expressed as a proportion of original population size for a given number of loci ($g$) under selection coefficient ($s$). See Figure 1 for conditions A – D.

Figure 3. Proportion of advantageous mutants in population ($f$) over time (generations) under mixed selection regimes of a single locus selected with $s$ of 0.3, and varying numbers of loci selected with $s$ of 0.01. A. Fixation of 0.3 selected locus in the presence of 0,1,5 and 10 loci at 0.01, a 0.3 locus in an initial population of 500 individuals, and a single 0.31 locus (out-crossing). B. A 0.01 selected locus in the presence of 0,1,5 and 10 other loci, one of which was selected at 0.3 and the others at 0.01 (out-crossing). C. A 0.3 selected locus in the presence of 0,1,5 and 10 loci at 0.01, a 0.3 locus in an initial population of 300 individuals, and a single 0.31 locus (inbreeding). D. A 0.01 selected locus in the presence of 0,1,5 and 10 other loci, one of which was selected at 0.3 and the others at 0.01 (inbreeding).

Figure 4. Proportion of advantageous mutants in population ($f$) over time (generations) with increasing numbers of loci in a uniform selection regime. A. 0.3 selected loci (outbreeding). B 0.01 selected loci (outbreeding). C. 0.3 selected loci (inbreeding). D. 0.01 selected loci (inbreeding).

Figure S1. Minimum population bottleneck (*b*) expressed as a percentage of original population size for a given number of loci (*g*). Loci selected at 0.005 for *mf* =1.5, selected at 0.01 for *mf* =10.

Figure S2. Probability of population survival for the number of loci under selection in inbreeding (blue) and outbreeding (red) systems from a standing frequency of the mutant under selection at 50%. A. selection coefficient *s* equal to 0.3. B. *s* equal to 0.01. C. *s* equal to 0.005. *mf* parameter set to 1.5.

Figure S3. Minimum population bottleneck expressed as a percentage of original population size for a given number of loci under selection I inbreeding (blue) and outbreeding (red) populations from a standing frequency of the mutant under selection at 50%. A selection coefficient *s* equal 0.3. . B. *s* equal to 0.01. C. *s* equal to 0.005. *mf* parameter set to 1.5.

Figure S4. Maximum selection load endured by populations without extinction for given values of *s*. A. In an *mf* =10 regime. B. In an *mf* = 1.5 regime.

Figure S5. The generations per substitution achieved by the time of estimated fixation under conditions of maximum selection load without extinction for given values of *s*.

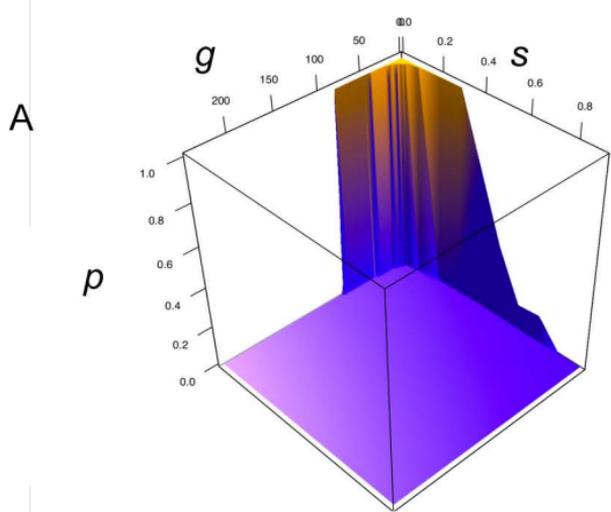

A

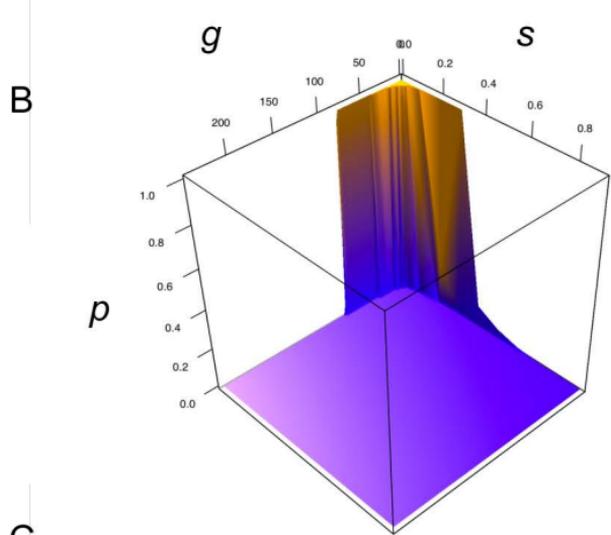

B

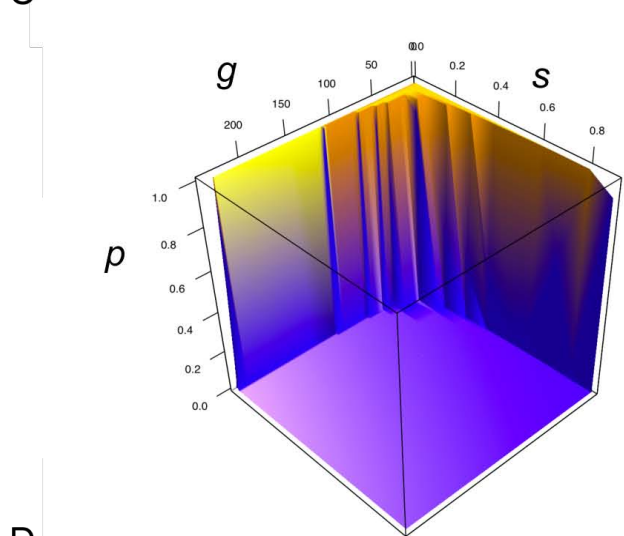

C

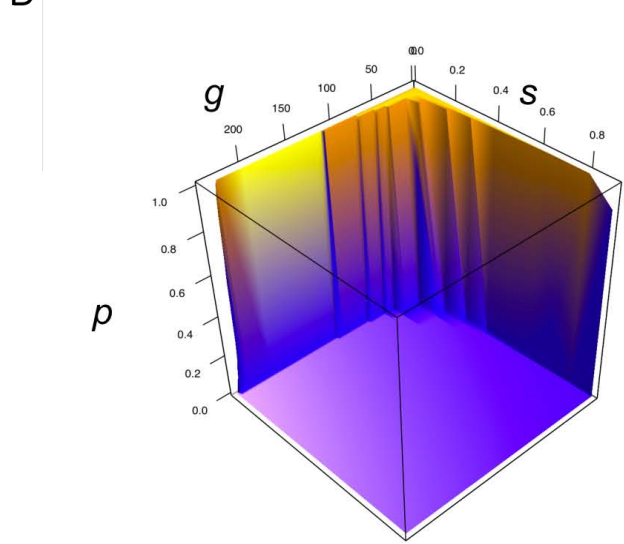

D

Figure 1

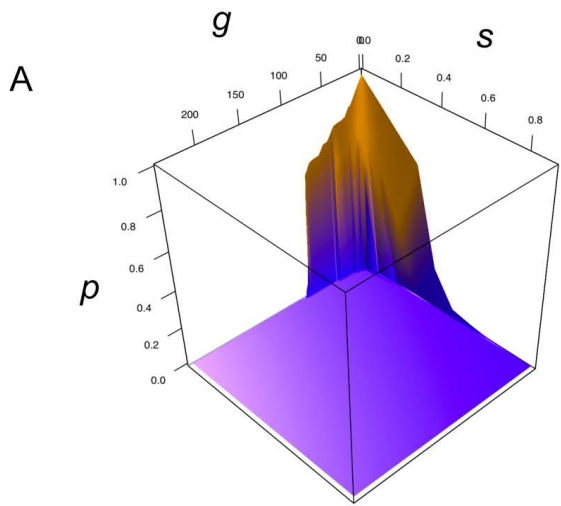

A

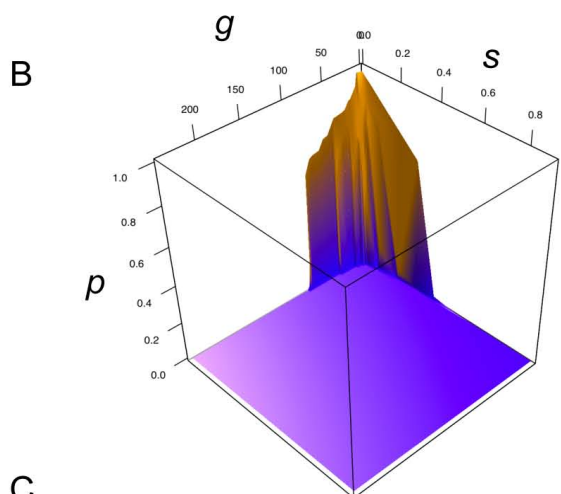

B

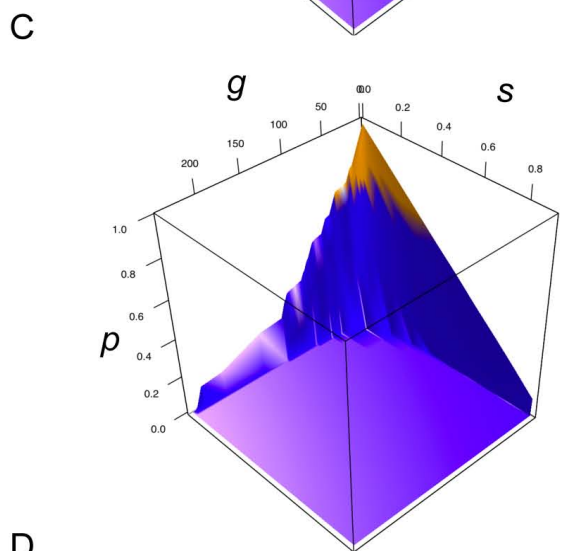

C

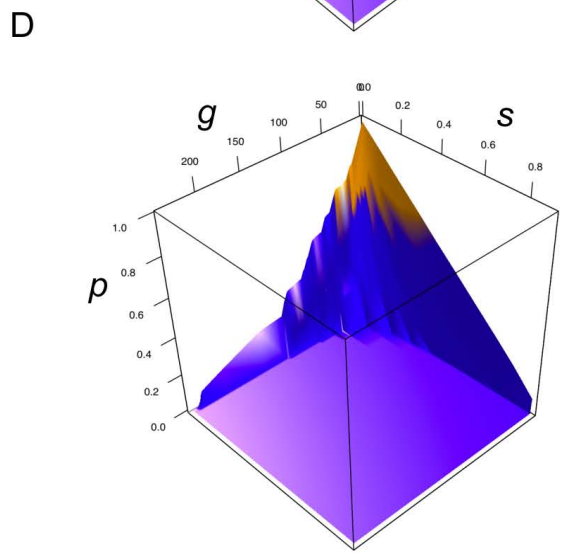

D

Figure 2

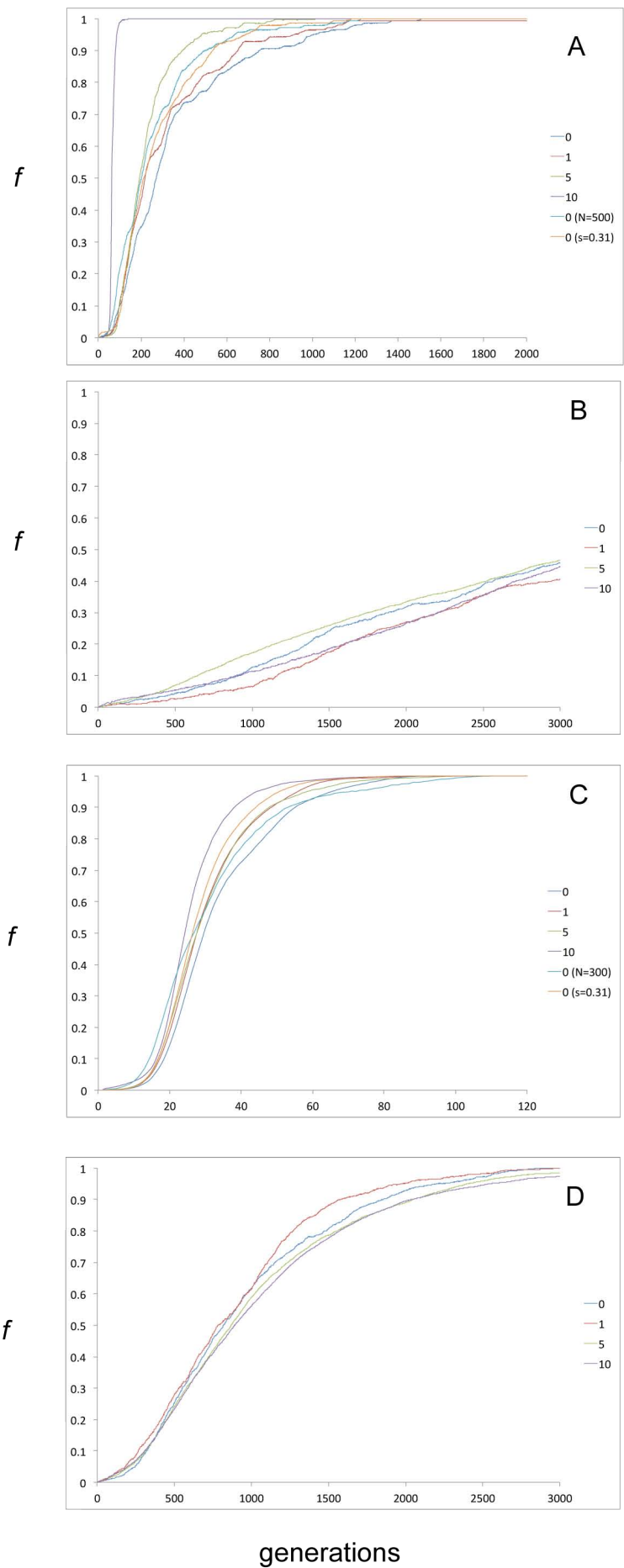

Figure 3

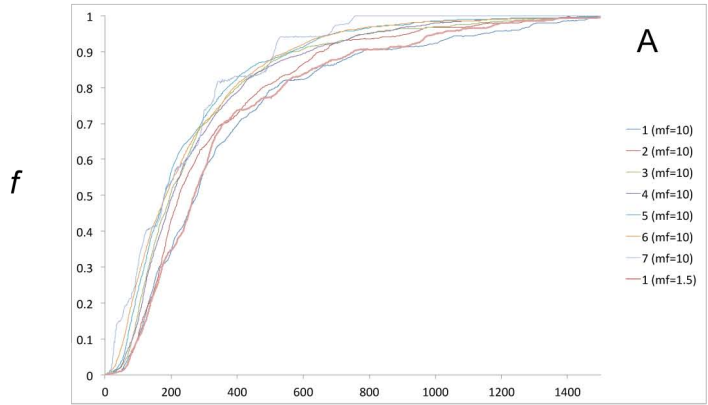
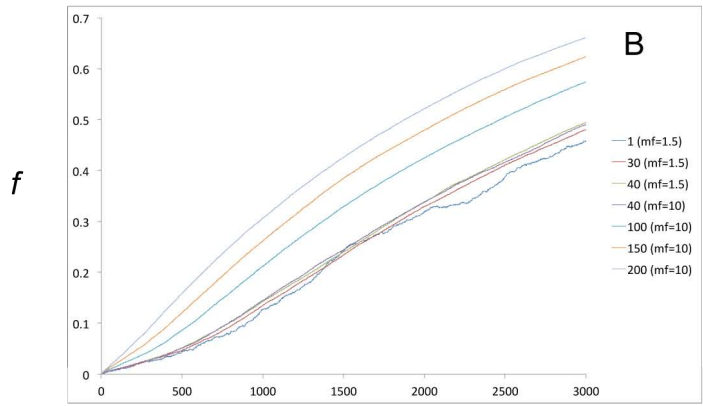
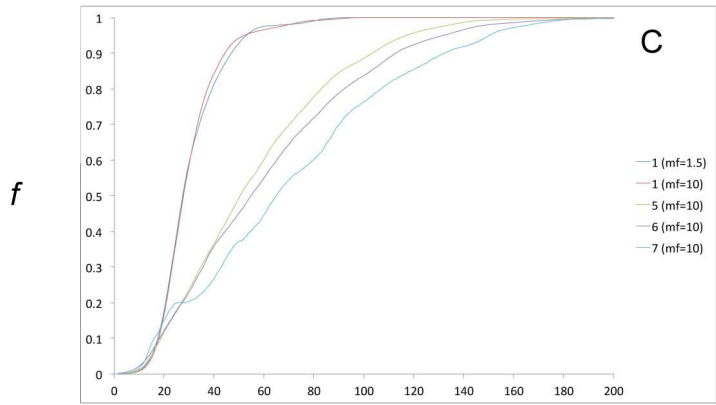
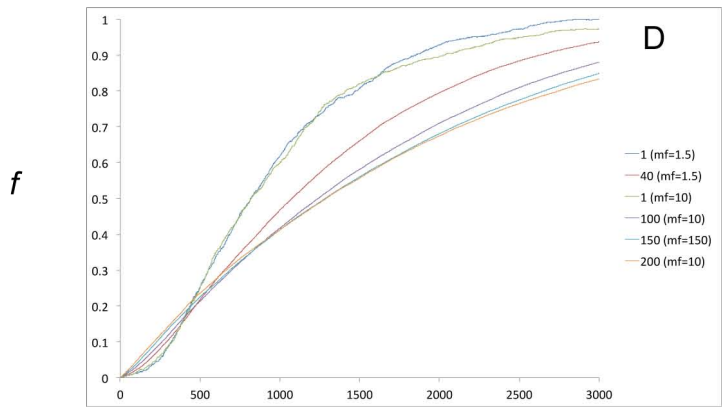

generations

Figure 4

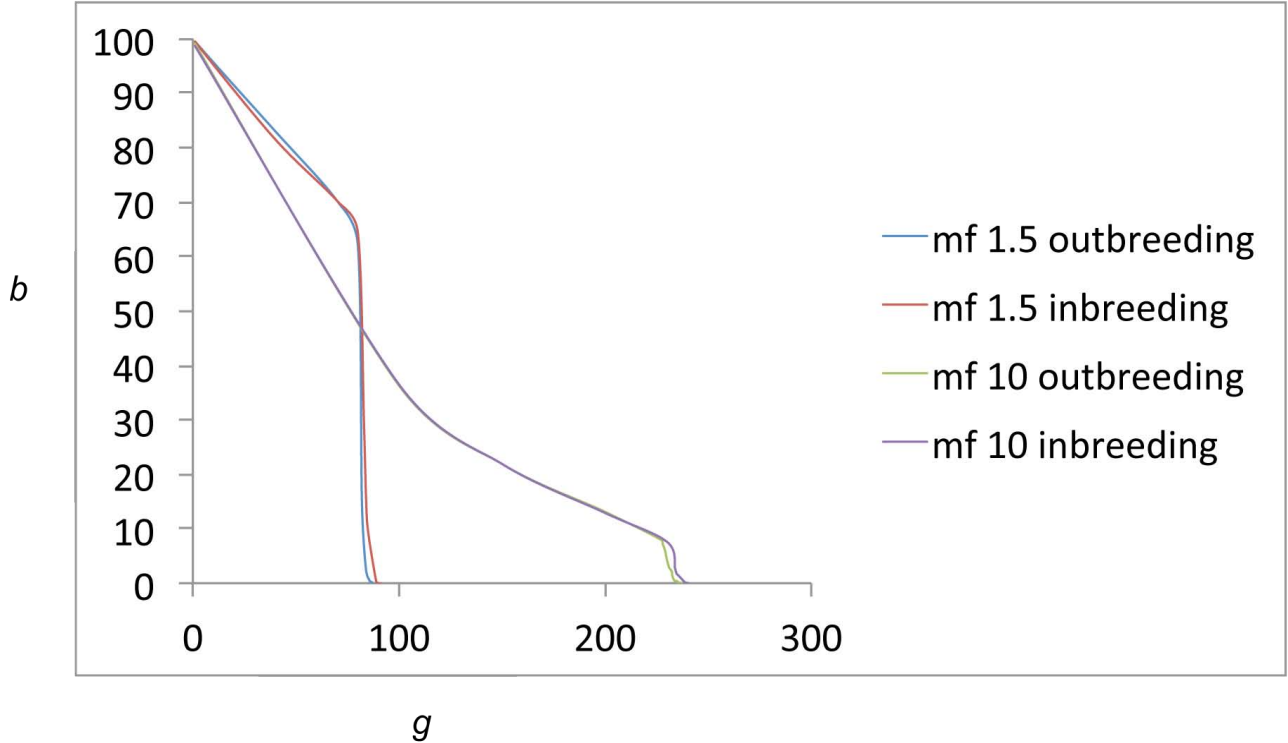

Figure S1

A

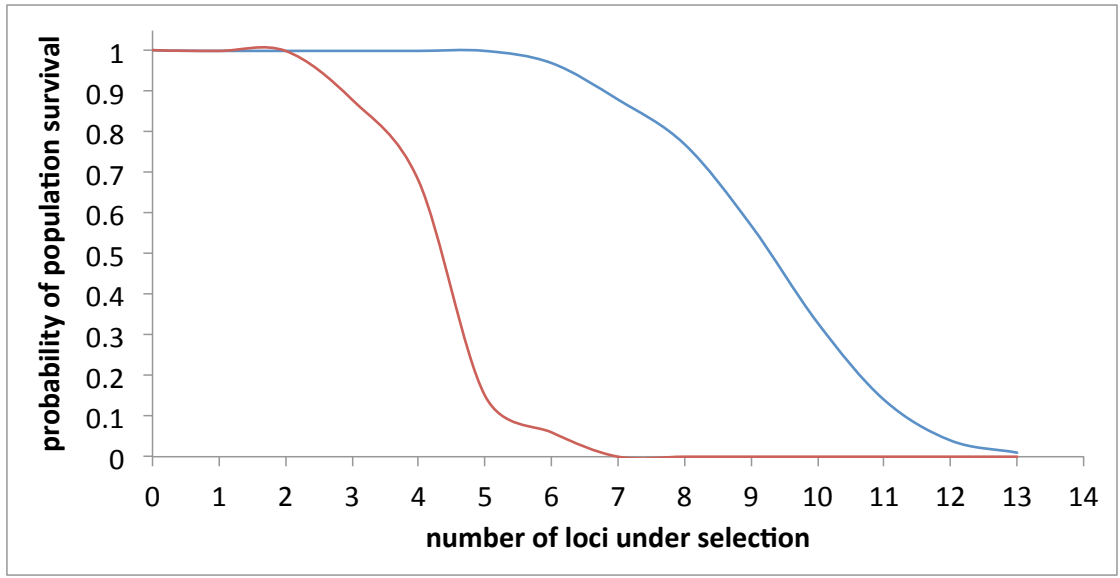

B

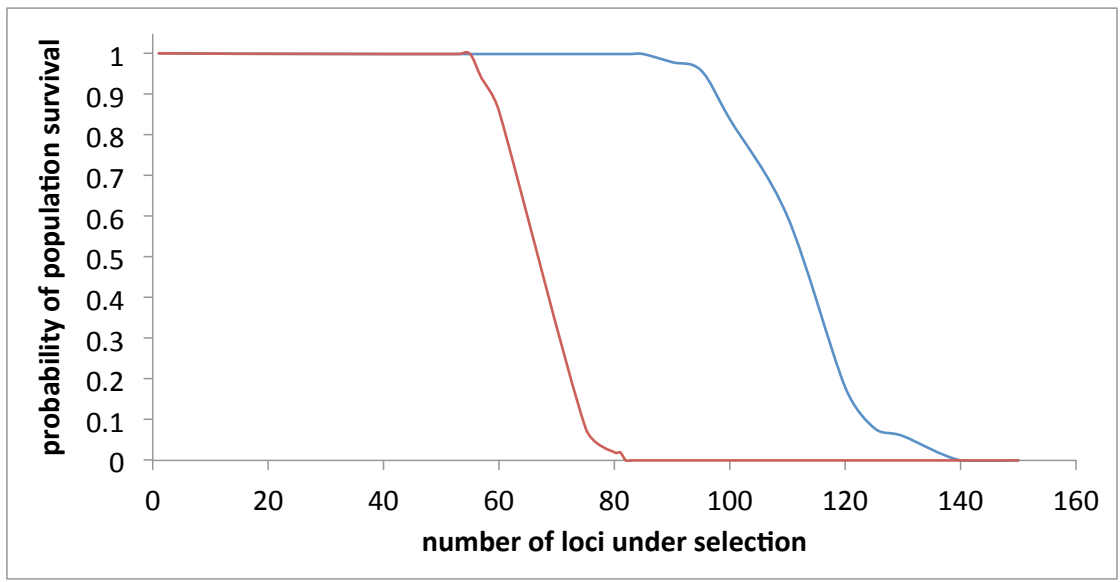

C

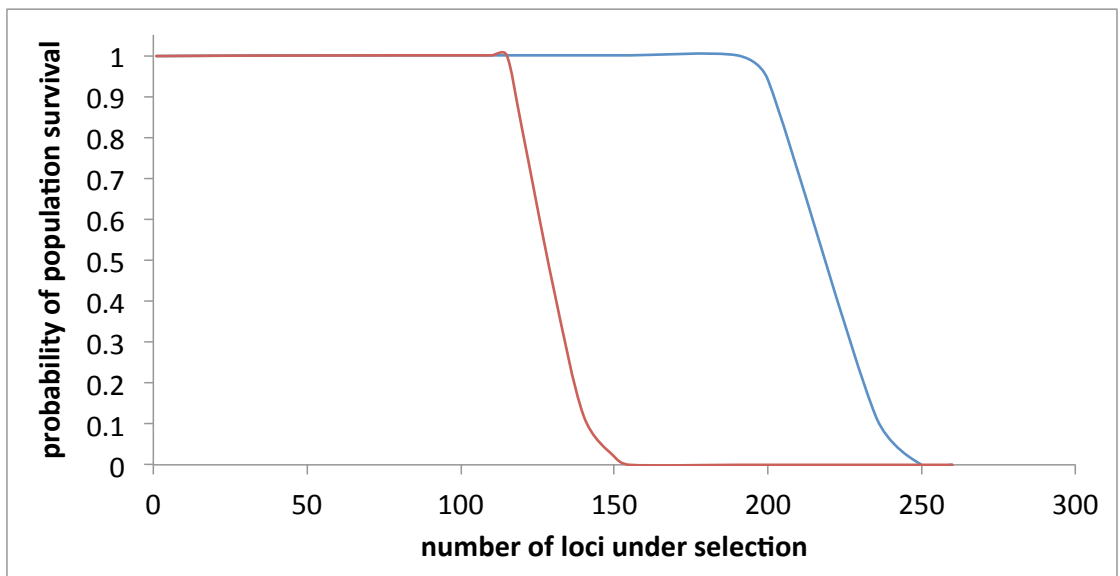

Figure S2

A

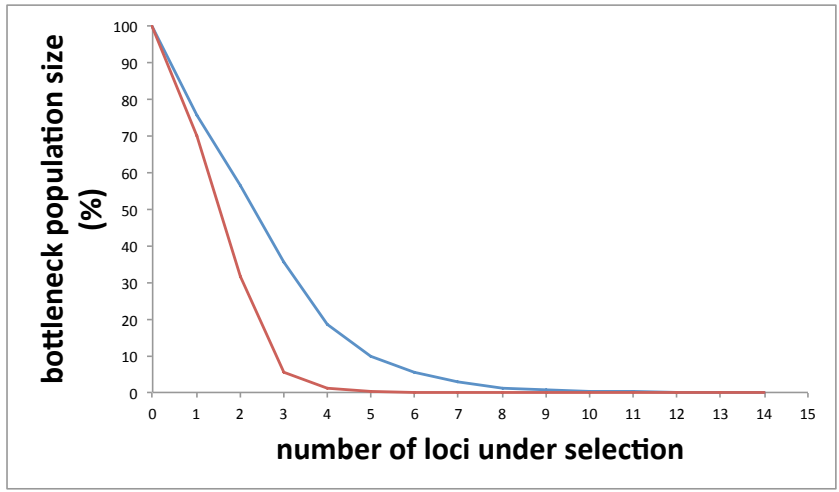

B

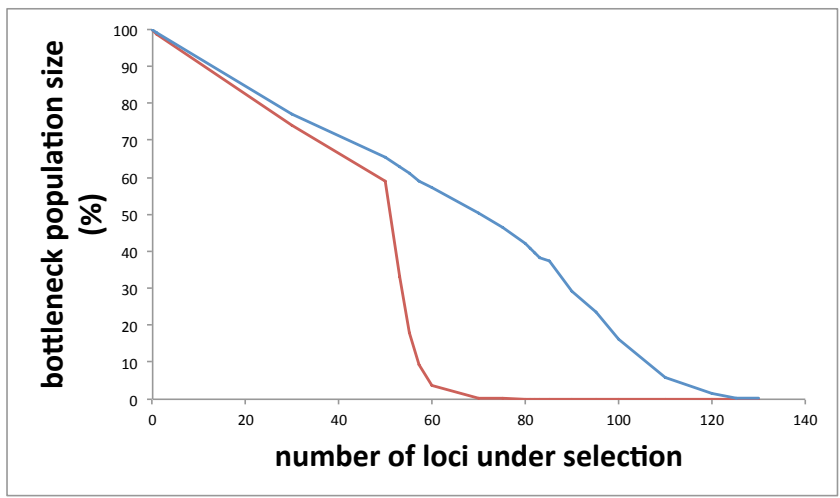

C

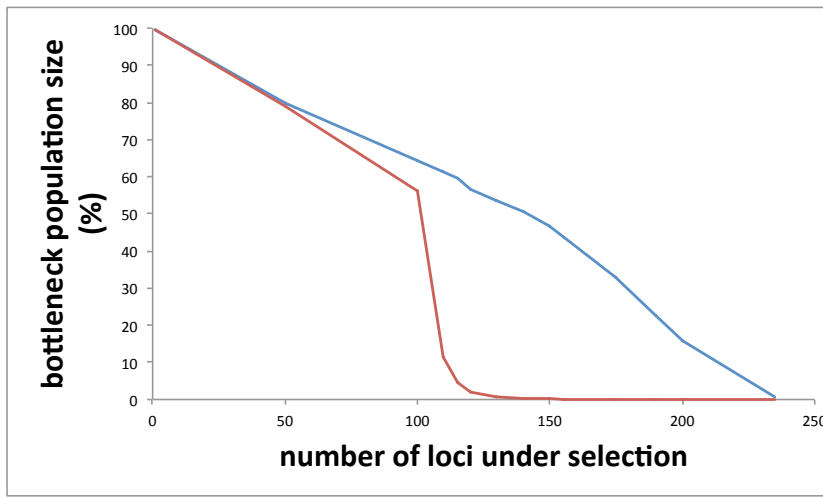

Figure S3

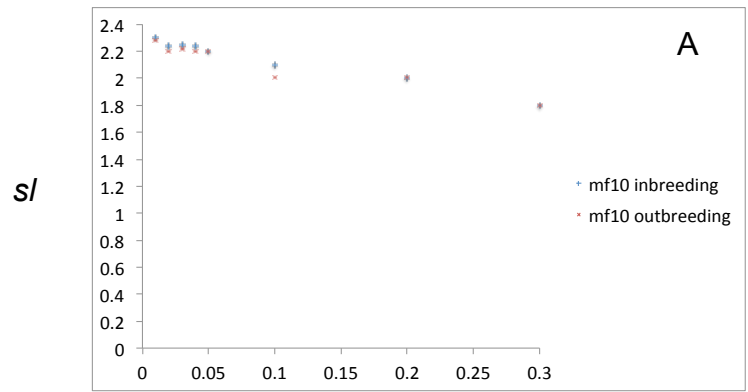
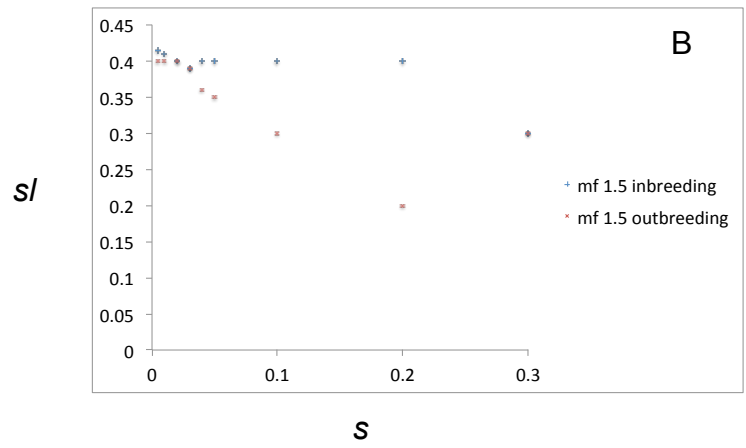

Figure S4

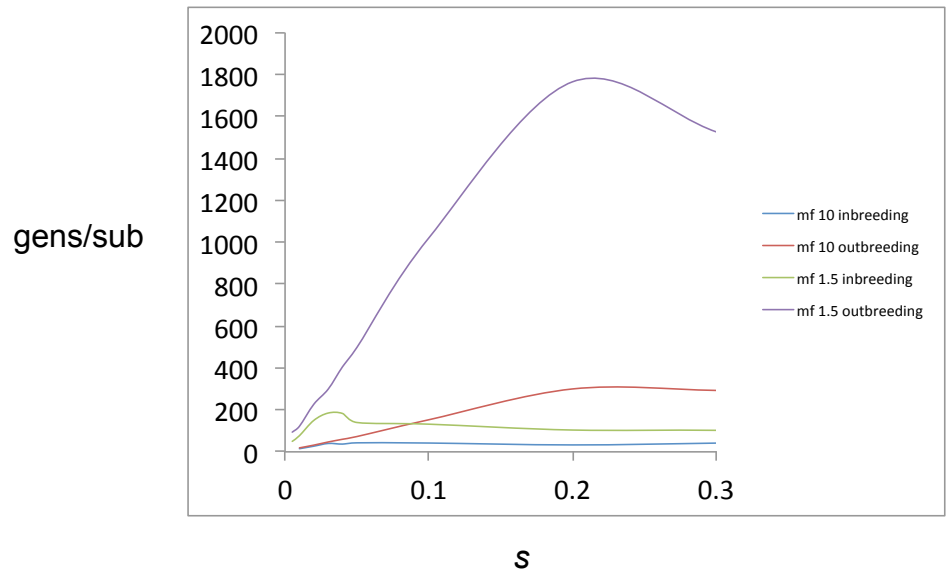

Figure S5

Table S1. Summary of simulation outputs for out-crossing populations with mf = 1.5

| s | loci number | %extinction | %frequency | %bottleneck |
|---|---|---|---|---|
| 0.9 | 1 | 100 | 0 | 0 |
| 0.9 | 2 | 100 | 0 | 0 |
| 0.8 | 1 | 100 | 0 | 0 |
| 0.8 | 2 | 100 | 0 | 0 |
| 0.7 | 1 | 100 | 0 | 0 |
| 0.6 | 1 | 99 | 99 | 0 |
| 0.6 | 2 | 100 | 0 | 0 |
| 0.5 | 1 | 96 | 96 | 0.125 |
| 0.5 | 2 | 100 | 0 | 0 |
| 0.4 | 1 | 90 | 90 | 1.774 |
| 0.4 | 2 | 100 | 100 | 0 |
| 0.4 | 3 | 100 | 100 | 0 |
| 0.3 | 1 | 0 | 0 | 69.525 |
| 0.3 | 2 | 98 | 98 | 0.481 |
| 0.3 | 3 | 100 | 100 | 0 |
| 0.2 | 1 | 0 | 100 | 79.634 |
| 0.2 | 2 | 81 | 100 | 3.996 |
| 0.2 | 3 | 100 | 100 | 0 |
| 0.2 | 4 | 100 | 100 | 0 |
| 0.1 | 1 | 0 | 98 | 89.696 |
| 0.1 | 2 | 0 | 98 | 80.549 |
| 0.1 | 3 | 0 | 98 | 72.646 |
| 0.1 | 4 | 42 | 99 | 12.865 |
| 0.1 | 5 | 100 | 100 | 0 |
| 0.05 | 1 | 0 | 85 | 94.9 |
| 0.05 | 2 | 0 | 89 | 90 |
| 0.05 | 3 | 0 | 86 | 85.5 |
| 0.05 | 4 | 0 | 89 | 81.2 |
| 0.05 | 5 | 0 | 86 | 77.1 |
| 0.05 | 6 | 0 | 85 | 73.108 |
| 0.05 | 7 | 0 | 86 | 69.569 |
| 0.05 | 8 | 11 | 88 | 29.523 |
| 0.05 | 10 | 100 | 0 | 0 |
| 0.04 | 1 | 0 | 75 | 95.8 |
| 0.04 | 2 | 0 | 79 | 91.8 |
| 0.04 | 3 | 0 | 77 | 88.2 |
| 0.04 | 4 | 0 | 81 | 84.7 |
| 0.04 | 5 | 0 | 79 | 81.3 |
| 0.04 | 10 | 7 | 84 | 34.048 |
| 0.04 | 11 | 95 | 83 | 0.33 |
| 0.04 | 12 | 100 | 0 | 0 |
| 0.03 | 1 | 0 | 75 | 96.7 |
| 0.03 | 2 | 0 | 70 | 93.8 |
| 0.03 | 3 | 0 | 72 | 91 |
| 0.03 | 4 | 0 | 72 | 88.24 |
| 0.03 | 5 | 0 | 73 | 85.571 |
| 0.03 | 10 | 0 | 77 | 73.4 |
| 0.03 | 12 | 0 | 75 | 69 |
| 0.03 | 13 | 0 | 77 | 66 |
| 0.03 | 14 | 61 | 79 | 3.5 |
| 0.03 | 15 | 96 | 86 | 0.009 |
| 0.03 | 16 | 100 | 0 | 0 |
| 0.03 | 20 | 100 | 0 | 0 |
| 0.02 | 1 | 0 | 59 | 97.7 |
| 0.02 | 10 | 0 | 63 | 81.4 |
| 0.02 | 20 | 0 | 66 | 53.3 |
| 0.02 | 21 | 65 | 70 | 2.8 |
| 0.02 | 22 | 98 | 84 | 0.01 |
| 0.02 | 23 | 100 | 0 | 0 |
| 0.02 | 30 | 100 | 0 | 0 |
| 0.01 | 1 | 0 | 51 | 98.8 |
| 0.01 | 10 | 0 | 45 | 90.1 |
| 0.01 | 20 | 0 | 46 | 81.3 |
| 0.01 | 30 | 0 | 48 | 73.7 |
| 0.01 | 40 | 0 | 49 | 61.8 |
| 0.01 | 41 | 13 | 54 | 14 |
| 0.01 | 42 | 59 | 55 | 2.9 |
| 0.01 | 43 | 96 | 57 | 0.2 |
| 0.01 | 45 | 100 | 0 | 0 |
| 0.005 | 1 | 0 | 26 | 99.3 |
| 0.005 | 40 | 0 | 32 | 83 |
| 0.005 | 70 | 0 | 33.8 | 70.2 |
| 0.005 | 80 | 0 | 35.5 | 62.2 |
| 0.005 | 82 | 5 | 40 | 14.8 |
| 0.005 | 84 | 62 | 42 | 2.4 |
| 0.005 | 86 | 95 | 46 | 0.3 |
| 0.005 | 88 | 100 | 0 | 0 |
| 0.005 | 90 | 100 | 0 | 0 |

Table S2. Summary of simulation outputs for inbreeding populations with mf = 1.5

| s | loci number | %extinction | %frequency | %bottleneck |
|---|---|---|---|---|
| 0.9 | 1 | 100 | 0 | 0 |
| 0.8 | 1 | 100 | 0 | 0 |
| 0.7 | 1 | 87 | 100 | 0.3 |
| 0.7 | 2 | 100 | 0 | 0 |
| 0.6 | 1 | 88 | 100 | 1.83 |
| 0.6 | 2 | 100 | 0 | 0 |
| 0.5 | 1 | 64 | 100 | 4.14 |
| 0.5 | 2 | 100 | 0 | 0 |
| 0.4 | 1 | 32 | 100 | 20.6 |
| 0.4 | 2 | 100 | 0 | 0 |
| 0.3 | 1 | 0 | 100 | 69.8 |
| 0.3 | 2 | 91 | 100 | 0.22 |
| 0.3 | 3 | 100 | 0 | 0 |
| 0.2 | 1 | 0 | 100 | 79.8 |
| 0.2 | 2 | 0 | 100 | 33.1 |
| 0.2 | 3 | 100 | 0 | 0 |
| 0.1 | 1 | 0 | 100 | 89.9 |
| 0.1 | 4 | 0 | 100 | 44.8 |
| 0.1 | 5 | 93 | 100 | 0.38 |
| 0.1 | 6 | 100 | 0 | 0 |
| 0.05 | 1 | 0 | 100 | 94.9 |
| 0.05 | 8 | 0 | 100 | 56.5 |
| 0.05 | 9 | 65 | 100 | 38.2 |
| 0.05 | 10 | 100 | 0 | 0 |
| 0.04 | 1 | 0 | 100 | 95.8 |
| 0.04 | 10 | 0 | 100 | 57.6 |
| 0.04 | 11 | 55 | 100 | 4.61 |
| 0.04 | 12 | 100 | 0 | 0 |
| 0.03 | 1 | 0 | 100 | 96.7 |
| 0.03 | 13 | 0 | 100 | 66 |
| 0.03 | 14 | 5 | 100 | 21.4 |
| 0.03 | 15 | 69 | 100 | 3.1 |
| 0.03 | 16 | 98 | 98 | 0.024 |
| 0.03 | 17 | 100 | 0 | 0 |
| 0.02 | 1 | 0 | 100 | 97.8 |
| 0.02 | 20 | 0 | 100 | 62 |
| 0.02 | 21 | 7 | 99 | 21.9 |
| 0.02 | 22 | 68 | 100 | 2.9 |
| 0.02 | 23 | 97 | 100 | 0.135 |
| 0.02 | 24 | 100 | 0 | 0 |
| 0.01 | 1 | 0 | 98 | 98.9 |

| 0.01  | 40 | 0   | 93   | 64    |
| 0.01  | 41 | 0   | 94   | 47    |
| 0.01  | 42 | 9   | 94   | 18    |
| 0.01  | 43 | 40  | 93.8 | 5.2   |
| 0.01  | 44 | 76  | 94.5 | 2.2   |
| 0.01  | 45 | 94  | 95   | 0.384 |
| 0.01  | 46 | 99  | 96.9 | 0.009 |
| 0.005 | 1  | 0   | 81.8 | 99.3  |
| 0.005 | 40 | 0   | 75   | 81.5  |
| 0.005 | 70 | 0   | 72   | 70.2  |
| 0.005 | 80 | 0   | 71   | 64.5  |
| 0.005 | 83 | 0   | 72.3 | 29.4  |
| 0.005 | 84 | 4   | 72   | 15.7  |
| 0.005 | 85 | 27  | 72.2 | 9.5   |
| 0.005 | 89 | 95  | 72.9 | 0.2   |
| 0.005 | 90 | 97  | 73.3 | 0.127 |
| 0.005 | 91 | 100 | 0    | 0     |

Table S3. Summary of simulation outputs for out-crossing populations with mf = 10.

| s | loci number | %extinction | %frequency | %bottleneck |
|---|---|---|---|---|
| 0.9 | 1 | 10 | 30 | 8.1 |
| 0.9 | 2 | 100 | 0 | 0 |
| 0.8 | 1 | 0 | 100 | 19.6 |
| 0.8 | 2 | 100 | 0 | 0 |
| 0.7 | 1 | 0 | 100 | 29.8 |
| 0.7 | 2 | 76 | 100 | 2.96 |
| 0.7 | 3 | 100 | 0 | 0 |
| 0.6 | 1 | 0 | 100 | 39.7 |
| 0.6 | 2 | 0 | 100 | 15.9 |
| 0.6 | 3 | 92 | 100 | 0.003 |
| 0.6 | 4 | 100 | 0 | 0 |
| 0.5 | 1 | 0 | 100 | 49.7 |
| 0.5 | 2 | 0 | 100 | 24.7 |
| 0.5 | 3 | 0 | 100 | 10.6 |
| 0.5 | 4 | 97 | 100 | 0.001 |
| 0.5 | 5 | 100 | 0 | 0 |
| 0.4 | 1 | 0 | 100 | 59.5 |
| 0.4 | 2 | 0 | 100 | 35.7 |
| 0.4 | 3 | 0 | 100 | 21.2 |
| 0.4 | 4 | 0 | 100 | 12.7 |
| 0.4 | 5 | 98 | 100 | 0.00148 |
| 0.4 | 6 | 100 | 0 | 0 |
| 0.3 | 1 | 0 | 100 | 69.7 |
| 0.3 | 2 | 0 | 100 | 48.5 |
| 0.3 | 3 | 0 | 100 | 34.1 |
| 0.3 | 4 | 0 | 100 | 23.7 |
| 0.3 | 5 | 0 | 100 | 16.7 |
| 0.3 | 6 | 0 | 100 | 11.56 |
| 0.3 | 7 | 96 | 100 | 0.00149 |
| 0.3 | 8 | 100 | 0 | 0 |
| 0.2 | 1 | 0 | 100 | 79.6 |
| 0.2 | 5 | 0 | 100 | 32.4 |
| 0.2 | 10 | 0 | 99.9 | 10.19 |
| 0.2 | 11 | 93 | 100 | 0.00285 |
| 0.2 | 12 | 100 | 0 | 0 |
| 0.1 | 1 | 0 | 94 | 89.78 |
| 0.1 | 5 | 0 | 98 | 58.69 |
| 0.1 | 10 | 0 | 97.8 | 34.61 |
| 0.1 | 15 | 0 | 98.3 | 20.26 |
| 0.1 | 20 | 0 | 97.7 | 11.86 |
| 0.1 | 22 | 32 | 98.69 | 3.76 |
| 0.1 | 23 | 95 | 98.38 | 0.167 |
| 0.1 | 24 | 100 | 0 | 0 |
| 0.05 | 1 | 0 | 87.41 | 94.8 |
| 0.05 | 20 | 0 | 91.53 | 35.8 |
| 0.05 | 30 | 0 | 91.5 | 21.3 |
| 0.05 | 44 | 0 | 92.7 | 9.4 |
| 0.05 | 45 | 19 | 92.5 | 5.1 |
| 0.05 | 46 | 79 | 92.8 | 0.75 |
| 0.05 | 47 | 100 | 0 | 0 |
| 0.04 | 1 | 0 | 85.5 | 95.8 |
| 0.04 | 20 | 0 | 86.2 | 44.03 |
| 0.04 | 40 | 0 | 88.6 | 19.3 |
| 0.04 | 55 | 0 | 90 | 9.7 |
| 0.04 | 56 | 2 | 90 | 7.5 |
| 0.04 | 57 | 49 | 90.3 | 2.76 |
| 0.04 | 58 | 92 | 90.7 | 0.00246 |
| 0.04 | 59 | 100 | 0 | 0 |
| 0.03 | 1 | 0 | 76.3 | 96.7 |
| 0.03 | 20 | 0 | 77.1 | 54.3 |
| 0.03 | 40 | 0 | 82.8 | 29.7 |
| 0.03 | 60 | 0 | 86.2 | 15.7 |
| 0.03 | 74 | 0 | 86 | 9.6 |
| 0.03 | 75 | 2 | 87 | 7.7 |
| 0.03 | 76 | 49 | 90.3 | 2.76 |
| 0.03 | 77 | 77 | 86.4 | 0.0092 |
| 0.03 | 78 | 95 | 84.8 | 0.00065 |
| 0.03 | 79 | | | |
| 0.02 | 1 | 0 | 69.1 | 97.8 |
| 0.02 | 40 | 0 | 71.3 | 44.2 |
| 0.02 | 60 | 0 | 74.7 | 29.3 |
| 0.02 | 80 | 0 | 77.2 | 19.74 |
| 0.02 | 100 | 0 | 79.1 | 13.1 |
| 0.02 | 110 | 0 | 79 | 10.5 |
| 0.02 | 112 | 0 | 79 | 9.5 |
| 0.02 | 113 | 0 | 79 | 8.4 |
| 0.02 | 115 | 53 | 79.4 | 2.1 |
| 0.02 | 116 | 77 | 80.4 | 0.7 |
| 0.02 | 117 | 84 | 79.1 | 0.4 |
| 0.02 | 118 | 94 | 81.4 | 0.19 |
| 0.02 | 119 | 100 | 0 | 0 |
| 0.01 | 1 | 0 | 39 | 98.9 |
| 0.01 | 100 | 0 | 57.4 | 36.3 |
| 0.01 | 150 | 0 | 62.3 | 21.9 |
| 0.01 | 200 | 0 | 66.1 | 13.1 |
| 0.01 | 227 | 0 | 67 | 8 |

| 0.01 | 228 | 2 | 67 | 7.1 |
| 0.01 | 229 | 6 | 67 | 6.1 |
| 0.01 | 230 | 14 | 67 | 4.3 |
| 0.01 | 231 | 34 | 67 | 2.9 |
| 0.01 | 232 | 44 | 67 | 2.3 |
| 0.01 | 233 | 74 | 68 | 0.88 |
| 0.01 | 234 | 84 | 69 | 0.58 |
| 0.01 | 235 | 86 | 68 | 0.32 |
| 0.01 | 236 | 92 | 69 | 0.18 |
| 0.01 | 237 | 94 | 68 | 0.082 |

Table S4. Summary of simulation outputs for inbreeding populations with mf = 10.

| s | loci number | %extinction | %frequency | %bottleneck |
|---|---|---|---|---|
| 0.9 | 1 | 7 | 100 | 9.24 |
| 0.9 | 2 | 100 | 0 | 0 |
| 0.8 | 1 | 0 | 100 | 19.8 |
| 0.8 | 2 | 95 | 100 | 0.19 |
| 0.8 | 3 | 100 | 0 | 0 |
| 0.7 | 1 | 0 | 100 | 29.9 |
| 0.7 | 2 | 48 | 100 | 6.1 |
| 0.7 | 3 | 100 | 0 | 0 |
| 0.6 | 1 | 0 | 100 | 39.7 |
| 0.6 | 2 | 0 | 100 | 15.8 |
| 0.6 | 3 | 85 | 100 | 0.4 |
| 0.6 | 4 | 100 | 0 | 0 |
| 0.5 | 1 | 0 | 100 | 50 |
| 0.5 | 2 | 0 | 100 | 24.8 |
| 0.5 | 3 | 0 | 100 | 12.7 |
| 0.5 | 4 | 88 | 100 | 0.039 |
| 0.5 | 5 | 100 | 0 | 0 |
| 0.4 | 1 | 0 | 100 | 59.7 |
| 0.4 | 4 | 0 | 100 | 12.9 |
| 0.4 | 5 | 75 | 100 | 1.06 |
| 0.4 | 6 | 100 | 0 | 0 |
| 0.3 | 1 | 0 | 100 | 69.8 |
| 0.3 | 5 | 0 | 100 | 16.7 |
| 0.3 | 6 | 0 | 100 | 11.6 |
| 0.3 | 7 | 75 | 100 | 1.07 |
| 0.3 | 8 | 100 | 0 | 0 |
| 0.2 | 1 | 0 | 100 | 79.8 |
| 0.2 | 10 | 0 | 100 | 10.4 |
| 0.2 | 11 | 66 | 100 | 1.37 |
| 0.2 | 12 | 100 | 0 | 0 |
| 0.1 | 1 | 0 | 100 | 89.9 |
| 0.1 | 21 | 0 | 100 | 10.6 |
| 0.1 | 22 | 6 | 100 | 7.5 |
| 0.1 | 23 | 90 | 100 | 0.58 |
| 0.1 | 24 | 99 | 100 | 0.04 |
| 0.1 | 25 | 100 | 0 | 0 |
| 0.05 | 1 | 0 | 100 | 94.8 |
| 0.05 | 45 | 0 | 100 | 7.6 |
| 0.05 | 46 | 44 | 100 | 2.5 |
| 0.05 | 47 | 91 | 100 | 0.5 |
| 0.05 | 48  | 97  | 100 | 0.008 |
| 0.05 | 49  | 100 | 0   | 0     |
| 0.04 | 1   | 0   | 100 | 95.9  |
| 0.04 | 56  | 0   | 100 | 8.9   |
| 0.04 | 57  | 10  | 100 | 5.6   |
| 0.04 | 58  | 64  | 100 | 1.6   |
| 0.04 | 59  | 87  | 100 | 0.5   |
| 0.04 | 60  | 99  | 100 | 0.004 |
| 0.04 | 61  | 100 | 0   | 0     |
| 0.03 | 1   | 0   | 100 | 96.8  |
| 0.03 | 75  | 0   | 100 | 8.6   |
| 0.03 | 76  | 6   | 100 | 6.8   |
| 0.03 | 77  | 34  | 100 | 3.6   |
| 0.03 | 78  | 73  | 100 | 1.37  |
| 0.03 | 79  | 90  | 100 | 0.4   |
| 0.03 | 80  | 100 | 0   | 0     |
| 0.02 | 1   | 0   | 100 | 97.8  |
| 0.02 | 112 | 0   | 98  | 9.5   |
| 0.02 | 113 | 1   | 98  | 9     |
| 0.02 | 115 | 14  | 98  | 5     |
| 0.02 | 117 | 63  | 99  | 1.8   |
| 0.02 | 118 | 81  | 98  | 0.6   |
| 0.02 | 119 | 91  | 98  | 0.4   |
| 0.02 | 120 | 100 | 0   | 0     |
| 0.01 | 1   | 0   | 97  | 98.8  |
| 0.01 | 100 | 0   | 88  | 36.5  |
| 0.01 | 150 | 0   | 85  | 21.9  |
| 0.01 | 200 | 0   | 83  | 12.9  |
| 0.01 | 230 | 0   | 83  | 7.6   |
| 0.01 | 234 | 48  | 83  | 2.6   |
| 0.01 | 235 | 58  | 82  | 1.8   |
| 0.01 | 238 | 86  | 81  | 0.4   |
| 0.01 | 239 | 96  | 82  | 0.1   |
| 0.01 | 240 | 98  | 86  | 0.06  |

Table S5. Summary of dual selection regime simulations

| loci number | | %extinction | %frequency | | %bottleneck | mating strategy |
|---|---|---|---|---|---|---|
| s=0.3 | s=0.01 | | 0.3 | 0.01 | | |
| 1 | 1 | 0 | 100 | 40 | 69 | out-crossing |
| 1 | 1 | 0 | 100 | 99.9 | 69 | inbreeding |
| 1 | 4 | 0 | 100 | 46 | 65 | out-crossing |
| 1 | 5 | 4 | 100 | 47 | 48 | out-crossing |
| 1 | 5 | 0 | 100 | 98 | 63 | inbreeding |
| 1 | 6 | 0 | 100 | 97 | 52 | inbreeding |
| 1 | 7 | 1 | 100 | 98 | 43 | inbreeding |
| 1 | 10 | 83 | 100 | 45 | 4 | out-crossing |
| 1 | 10 | 22 | 100 | 97 | 32 | inbreeding |
| 1 | 19 | 99 | 100 | 54 | 0.06 | out-crossing |
| 1 | 19 | 66 | 100 | 48 | 7 | inbreeding |

Supplementary Information

**Population selection cost program methodology**

The program used to simulate selection cost was written in perl by RG Allaby, available to download from the Allaby group website: (http://www2.warwick.ac.uk/fac/sci/lifesci/research/archaeobotany/downloads/).

*Initialization*
The program consists of the core algorithm (population_selector.pl), and a wrap that determines the number of times the core algorithm is executed, and extracts summary statistics (selector.pl). The input parameters are contained in a text file (parameters). A second text file (selection_coefficients) lists the selection coefficients and values of lambda for each locus, if there are variable values between loci.

The parameters the program takes are as follows (grey indicates parameter values not used in this study, but require an initializing value of 0 input):

1. population size
2. number of genes under selection
3. selection coefficient (if there is just one)
4. maximum fecundity (see main manuscript for an explanation of this)
5. number of generations to simulate
6. number of sexes - 1 is hemaphrodite, 2 is males and females
7. unused variable
8. dominance options (a) 1 all have same lambda, 2 randomly assign lambda to different mutants
9. dominance options (b)lambda (if all the same) determines dominance, 0 fully recessive, 1 fully dominant.
10. unused variable
11. mutation rate
12. variable selection coefficient option: 0 all have same s value determined by option 3, 1 look up selection coefficients and associated lambda values in selection_coefficients file.
13. mating strategy, 0-1, the proportion of matings that are self-fertilization events.
14. gene flow, 0-1 immigrant rate as defined by a proportion of the carrying capacity of individuals arriving from a non-selective environment each generation

*The core algorithm (population_selection.pl)*
The simulation begins with an initialization of the population in which individuals are assigned wild type alleles for all loci under selection. A log of wild type frequency and mutant frequency is written to an output file (outfile) for each generation, as well as the population size for that generation. An advantageous mutation occurred with a probability equal to the mutation rate, and was attributed to an individual in the population. The next generation of individuals was then generated by randomly selecting an individual from the previous generation, from which a gamete would be generated by randomly selecting one or other of the alleles of that individual for each locus. Therefore all loci are modeled as unlinked in this simulation. A second gamete was then generated with a probability equal to the mating strategy variable of selecting the same parent source. Once an individual had been generated, it was challenged such that its probability of survival was given as (*su*):

$$su = \prod_{1}^{k} \omega_i \qquad (1)$$

For *k* loci, where $\omega_I$ is the fitness of the *i*th locus as given by:

$$\omega_i = 1 - s_i \qquad (2)$$

Where $s_i$ is the selection coefficient of the *i*th locus. The selection coefficient of the *i*th locus was moderated by the value lambda for heterozygotes ($S_{het}$) such that

$$s_{ihet} = \lambda s_i \qquad (3)$$

A value of 0 was taken for lambda in all simulations in this study to represent recessive mutations, which represent the majority of known mutations associated with domestication.

In the first generation this step was repeated for a number of times equal to the population size, and inevitably led to a number of individuals in the next generation which were fewer than this value. In subsequent generations the number of attempts at making new individuals was given by

$$N_{attempts} = N_{n-1} mf \qquad (4)$$

For ($N_{attempts}$ < initial population size), where $N_{attempts}$ is the number of individuals created then challenged, $N_{n-1}$ is the number of individuals in the previous generation and *mf* is the maximum fecundity parameter. Where this condition was violated, the initial population size was used, representing the carrying capacity of the environment. This process was iterated for the specified number of generations.

### *The wrap algorithm (selector.pl)*
The second program calls the first core algorithm for a defined number of trials and in each trial records the out put file (outfile), and calculates the mean and variance each generation of the population size, and the mutant allele frequency for each locus. The mean of the means are then calculated for all loci for each generation, as well as the number of trials in which populations went extinct each generation and the cumulative proportion of trials in which extinction had occurred. These values are collated into a second out put file (otheroutfile).

### *Standing variation*
A variant of the population_selector.pl program was used to study standing variation called population_selector_SV.pl. This required some changes in data structure in initialize populations to account for standing variation which were set (hard coded) to start at frequencies of 50% in genotype proportions that accurately reflected the mating strategy.